\documentclass[12pt,preprint]{aastex}

\def\paren#1{\left( #1 \right)}

\shorttitle{GRB~061007: Optical Flashes and GRB Fireballs}
\shortauthors{Mundell et al.}

\begin{document}


\title{The Remarkable Afterglow of GRB~061007:
Implications for Optical Flashes and GRB Fireballs}

\author{C.~G.~Mundell\altaffilmark{1}, A.~Melandri\altaffilmark{1},
C.~Guidorzi\altaffilmark{1,2,3},  S.~Kobayashi\altaffilmark{1},
I.~A.~Steele\altaffilmark{1}, D.~Malesani\altaffilmark{4},
L.~Amati\altaffilmark{5}, P.~D'Avanzo\altaffilmark{3,6}, D.~F.~Bersier\altaffilmark{1}, A.~Gomboc\altaffilmark{1,7},
 E.~Rol\altaffilmark{8}, M.~F.~Bode\altaffilmark{1}, D.~Carter\altaffilmark{1},  C.~J.~Mottram\altaffilmark{1},  A.~Monfardini\altaffilmark{1,9}, R.~J.~Smith\altaffilmark{1}, S.~Malhotra\altaffilmark{10}, J.~Wang\altaffilmark{11}, N.~Bannister\altaffilmark{8},  P.~T.~O'Brien\altaffilmark{8},
 N.~R.~Tanvir\altaffilmark{8}.}

\email{(cgm)@astro.livjm.ac.uk}
\altaffiltext{1}{Astrophysics Research Institute, Liverpool John Moores
University, Twelve Quays House, Birkenhead, CH41 1LD, UK}
\altaffiltext{2}{Dipartimento di Fisica, Universit\'a di Milano-Bicocca, Piazza 
delle Scienze 3, 20126 Milano, Italy}
\altaffiltext{3}{INAF-Osservatorio Astronomico di Brera, via Bianchi 46, 23807 Merate (LC), Italy}
\altaffiltext{4}{International School for Advanced Studies (SISSA-ISAS), via Beirut 2-4, I-34014 Trieste, Italy}
\altaffiltext{5}{INAF - IASF Bologna, via P. Gobetti 101, Bologna, Italy}
\altaffiltext{6}{Universit\`a dell'Insubria, Dipartimento di Fisica e Matematica, via 
Valleggio 11, I-22100 Como, Italy}
\altaffiltext{7}{FMF, University in Ljubljana, Jadranska 19, 1000
Ljubljana, Slovenia.}
\altaffiltext{8}{Department of Physics and Astronomy, University of Leicester, University Road, Leicester LE1 7RH, UK}
\altaffiltext{9}{CNRS-CTBT, 25 Avenue des Martyrs, 38000 Grenoble, France}
\altaffiltext{10}{School of Earth and Space Exploration, Arizona State University, P.O. Box 871404, Tempe, AZ 85287-1404}
\altaffiltext{11}{Center for Astrophysics, University of Science and Technology of China, Hefei, Anhui 230026, P. R. China}


\begin{abstract} We present a multiwavelength analysis of {\em Swift}
GRB~061007. The 2-m robotic Faulkes Telescope South (FTS) began
observing 137~s after the onset of the $\gamma$-ray emission, when the
optical counterpart was already decaying from R$\sim$10.3~mag, and
continued observing for the next 5.5 hours. These observations begin
during the final $\gamma$-ray flare and continue through and beyond a
long, soft tail of $\gamma$-ray emission whose flux shows an
underlying simple power-law decay identical to that seen at optical
and X-ray wavelengths, with temporal slope $\alpha\sim1.7$
($F\propto~t^{-\alpha}$).  This remarkably simple decay in {\em all}
of these
 bands is rare for {\em Swift} bursts, which often show much more
 complex light curves.  We suggest the afterglow emission begins as
 early as $30-100$ s and is contemporaneous with the on-going variable
 prompt emission from the central engine, but originates from a
 physically distinct region dominated by the forward shock. The
 afterglow continues unabated until at least $\sim$10$^5$ seconds
 showing no evidence of a break.  The observed multiwavelength
 evolution of GRB~061007 is explained by an expanding fireball whose
 optical, X-ray and late-time $\gamma$-ray emission is dominated by
 emission from a forward shock with typical synchrotron frequency,
 $\nu_m$, that is already below the optical band as early as t=137~s
 and a cooling frequency, $\nu_c$, above the X-ray band to at least
 t=10$^5$~s. In contrast, the typical frequency of the reverse shock
 lies in the radio band at early time.  We suggest that the unexpected
 lack of bright optical flashes from the majority of {\em Swift} GRBs
 may be explained with a low $\nu_m$ originating from small microphysics
 parameters, $\epsilon_e$ and $\epsilon_B$. We derive a minimum jet
 opening angle $\theta=4.7^{\circ}$ from the optical light curves and
conclude that GRB 061007 is a  secure outlier to spectral energy
correlations because no X-ray jet break occurred before $t\sim10^6$ s.

\end{abstract}


\section{Introduction} Gamma-ray bursts (GRBs) are the most
instantaneously luminous cosmological objects known, and although
their transient nature makes them observationally challenging to
follow up at other wavelengths after the initial burst of $\gamma$
rays, their brightness and large redshift range ($0.0085<z<6.29$)
ensure they remain uniquely useful as probes of the early Universe.
With the launch of the GRB-dedicated {\em Swift} satellite
\citep{Gehrels04}, early-time monitoring of the radiation from GRBs
across the electromagnetic spectrum is now possible. Rapid and
accurate localisation of GRBs and immediate dissemination of their
positions drive ground-based follow-up campaigns with small and large
optical and infrared robotic telescopes such as ROTSE, REM, the
Liverpool and Faulkes Telescopes, while Swift's onboard narrow field
instruments the X-Ray Telescope (XRT) and the UltraViolet Optical
Telescope (UVOT) observe a large fraction of GRBs within minutes of
their detection with the Burst Alert Telescope (BAT).

Following the detection of a very bright optical counterpart to
GRB~990123 by ROTSE \citep{Akerlof99} contemporaneous with the
$\gamma$-ray emission and consistent with models predicting optical
emission from a reverse shock in the relativistic
outflow~\citep{Sari99,Kobayashi00a,Kobayashi00b,Nakar05}, there was
strong anticipation of routine detection of optical flashes at early
times with UVOT, which can detect optical emission below the detection
threshold of ROTSE. However, despite rapid and accurate localisation
of $\gamma$-ray emission and subsequent follow-up by UVOT, bright
optical flashes remain elusive (Roming et al. 2006). In some cases,
the high redshift of the burst renders it undetectable due to the lack
of UVOT filters redder than V-band (i.e. z$>$4); in others, intrinsic
faintness coupled with local extinction pushes the OT below the
detection threshold of the UVOT~\citep{Oates06}. Ground-based
observations from large robotic telescopes, optimised for the rapid
follow-up of GRB alerts~\citep{Guidorzi05,Guidorzi06a,Dai06}, are
increasing the number of well-sampled, early-time multicolor optical
light curves over a wide range of magnitudes (Gomboc et al. 2006;
Melandri et al. 2006) and revealing more complex decays than expected
from pre-Swift light curves (Guidorzi et al. 2005; Stanek et al. 2006;
Dai et al. 2006; Monfardini et al. 2006; Oates et al. 2006). Small
departures from a power-law decay (i.e. $\Delta$m$\lesssim$0.5 mag)
are interpreted as interactions between the expanding fireball and
dense clumps in the circumburst medium (Lazzati et al. 2002; Heyl \&
Perna 2003; Nakar et al. 2003; Guidorzi et al. 2005) while major
rebrightenings suggest late-time energy injections (Kumar \& Piran
2000; Sari \& M\'esz\'aros 2000; Zhang \& M\'esz\'aros 2002; Stanek et
al. 2006; Wozniak et al. 2006; Monfardini et al. 2006); bright optical
flares consistent with reverse shocks, however, are still rare or
unconfirmed (Bo\"er et al. 2006; Jel\'{\i}nek et al. 2006).  Complex
light curves are now common at X-ray energies with {\em Swift} light
curves revealing a canonical shape comprising a ``3-segment'' structure
inconsistent with a single power-law decay, with X-ray flares
superimposed in at least 50\% of bursts. Such complexity has been a
major surprise of the {\em Swift} era and has led to a general
acceptance of long-lived activity~\citep{Burrows05,Obrien06},
challenging standard central engine models. A significant minority of
GRB X-ray light curves show a fairly smooth decay consistent with an
afterglow at early time (O'Brien et al. 2006) but most of these cases
lack simultaneous optical data sufficient to fully test current models.

Here we present multicolor BVR$i^\prime$ light curves of the very
bright optical counterpart to GRB~061007 obtained automatically by the
2-m robotic telescope, the Faulkes Telescope South (FTS) at Siding
Spring, Australia which began observing 137~s after the onset of the
$\gamma$-ray emission when the optical counterpart was already
decaying from R$\sim$10.3~mag, and continued observing for the next
5.5 hours\footnote{Observations obtained as part of the RoboNet-1.0
project: http://www.astro.livjm.ac.uk/RoboNet/}. We show that
GRB~061007 is unusual for a {\em Swift} burst in that it shows a
remarkably simple power-law decay from $\gamma$ rays to optical
wavelengths. Its isotropic equivalent radiated energy is one of the
highest ever measured (E$_{\rm iso}\sim10^{54}$ erg) and its peak
$\nu F_{\nu}$ energy, E$_{\rm peak}$, is well-determined. 
It therefore provides a useful test of current spectral
energy correlations without the complication of highly-structured
light curves~\citep{Amati02,Ghirl04,Liang05,Firmani06,Amati06}.

The multiwavelength observations and analyses are presented in
Section 2; results and discussion are presented in Section 3, where a
shock model is proposed to explain the observed multiwavelength
evolution and broad-band spectral energy distribution. We end by
considering implications for cosmological spectral energy correlations
and present our conclusions in Section 4. Throughout, 1-$\sigma$ errors are given unless otherwise stated.

\section{Observations and Analysis}
\label{sec:obs}

On 2006 Oct 07, 10:08:08 UT\footnote{t=0 throughout the paper.}  {\em
Swift}-BAT detected GRB~061007 (trigger 232683) at coordinates
R.A.(J2000)=03:05:16, Dec(J2000)=$-$50:29:15 (3 arc-min 90\%
containment radius). The prompt emission from GRB~061007 was also
detected at MeV energies by Konus/{\em Wind} (20~keV$-$10~MeV)
(Golenetskii et al. 2006), {\em Suzaku}/WAM (50~keV$-$5~Mev) (Yamaoka
et al. 2006) and {\em RHESSI} (20~keV$-$10~MeV) (Wigger et al. 2006),
whose detections provide estimates of the peak energy E$_{\rm peak}$
of 399$_{-18}^{+19}$ keV (Golenetskii et al. 2006), 561$_{-27}^{+29}$
keV (Yamaoka et al. 2006) and 391$_{-52}^{+58}$ keV (Wigger et al.
2006) respectively.  The {\em Swift}-BAT $\gamma$-ray light curve
shows three significant flat-topped peaks with substantial
sub-structure, and a small fourth peak (from t0+75s) that shows long
exponential decay and faint emission detectable to
t$\sim$~900~s~\citep{Markwardt06,Schady06c}. The fluence,
S$_{\gamma}$=2.49$\times$10$^{-4}$~erg~cm$^{-2}$, measured by
Konus-Wind, corresponds to an isotropic equivalent energy E$_{\rm
iso}$$\sim$10$^{54}$~erg~\citep{Golen06} for the spectroscopic
redshift z=1.261 (Osip et al. 2006; Jakobsson et al. 2006). The {\em
Swift} XRT and UVOT began observing 80~s after the BAT trigger time and
identified a very bright X-ray/optical counterpart with a rapid decay
rate $\alpha_x$=1.6$\pm$0.1 (F~$\propto$~t$^{-\alpha}$)
\citep{Schady06a,Vetere06,Schady06b}.

Robotically-triggered photometric observations with ground-based
telescopes ROTSE and the FTS began at 26 and 137 s respectively, each
identifying the optical counterpart with brightnesses 13.6~mag
(unfiltered)~\citep{Rykoff06} and 10.3~mag (R-band) (Mundell et
al. 2006; Bersier et al. 2006) respectively at
R.A.(J2000)=03:05:19.56, Dec(J2000)=$-$50:30:2.6
($\pm$1\farcs2). After taking the initial $3\times10$~s exposure
R-band images, the FTS continued to observe for the next 5.5 hours,
cycling through BR$i^\prime$ filters with V imaging interspersed and
using increasing exposure times, until finishing with a 300-s V-band
image at UT=15:39:25.322. Initial photometric calibration was
performed with GAIA (Optimal PSF Photometry Tool) relative to all the
catalogued stars (USNO B1 and NOMAD) in the 4\farcm6$\times$4\farcm6
field of view and the bright star near the OT (RA=03:05:23.5,
Dec=-50:30:16.6) was used for PSF determination.  A $5\times120$~s
R-band image was also acquired using FORS1 on the VLT at 2006~Oct
8.04702 UT (0.6247 days after the burst); the observing conditions
were photometric and calibration was performed relative to Landolt
standard stars.  The FTS R-band were adjusted to the VLT photometry to
provide a well-calibrated R-band light curve. Later-time R-band images
were acquired with the IMACS instrument on the Magellan-I Baade
telescope at Las Campanas Observatory on Oct 9 and 10 and calibrated
with respect to the VLT image. Finally, magnitudes are corrected for
Galactic extinction: from $E_{B-V}=0.020$ \citep{Schlegel98}, and
$A_V=R_V\cdot E_{B-V}=0.06$ mag, with $R_V=3.1$, we evaluated the
extinction in the other filters following Cardelli et al. (1989):
$A_B=0.08$ mag, $A_{R}=0.05$ mag and $A_{i^{\prime}}=0.04$ mag.  Fluxes
were derived following Fukugita et al. (1996) and calibration
uncertainties for the remaining FTS filters, which do not affect the
overall light curve properties, were taken into account when deriving
the broad-band spectral energy distribution. Results are summarized in
Table~\ref{tab:obsfts} and plotted in
 Figure~\ref{fig:LC} along with the mask-tagged background subtracted
 BAT light curve, which was produced by the standard BAT pipeline
 (HEADAS v 6.1.1).  The BAT light curve was masked with the
 ground-refined coordinates provided by the BAT team
 \citep{Markwardt06}. BAT spectra have been corrected for systematics
 with the ``{\em batphasyserr}'' tool and the original 80 energy
 channels have been grouped to have 3-$\sigma$ significant grouped
 channels.  We processed XRT data from 200~s to 400~s after the
 trigger adopting standard screening and we extracted the XRT
 0.3--10~keV spectrum: data were in Window Timing (WT) mode and the
 average count rate was below 100~counts~s$^{-1}$ so that no
 correction for pileup was required \citep{Romano06}.  

The brightness of the burst and the similarity of the decay rate in
all bands (see Figure~\ref{fig:LC} and Table~\ref{tab:alphas}) allows
the construction of an early-time broad-band spectral energy
distribution (SED), shown in Figure~\ref{fig:SED}.  The joint XRT/BAT
0.3--150~keV spectrum from 200~s to 400~s is well fit with a power law
with significant absorption in excess of the Galactic, $N_{\rm
H}(Gal)=1.75\times10^{20}$~cm$^{-2}$ \citep{Kalberla05}, such that
$\beta_{{\rm X-}\gamma}=1.01\pm0.03$ (F~$\propto\nu^{-\beta}$) and
rest-frame $N_{\rm H}=(5.8\pm0.4)\times10^{21}$~cm$^{-2}$ (90\%
confidence level). As shown in Figure~\ref{fig:SED} and
Table~\ref{tab:sedfits}, when $\beta$ is free to vary, the best fit is
obtained for $\beta_{{\rm opt-X-}\gamma}=1.02\pm0.05$ and
$A_V$=0.48$\pm$0.19 mag with an SMC-like extinction profile \citep{Pei92}.

\section{Results and Discussion}

\begin{figure}
\plottwo{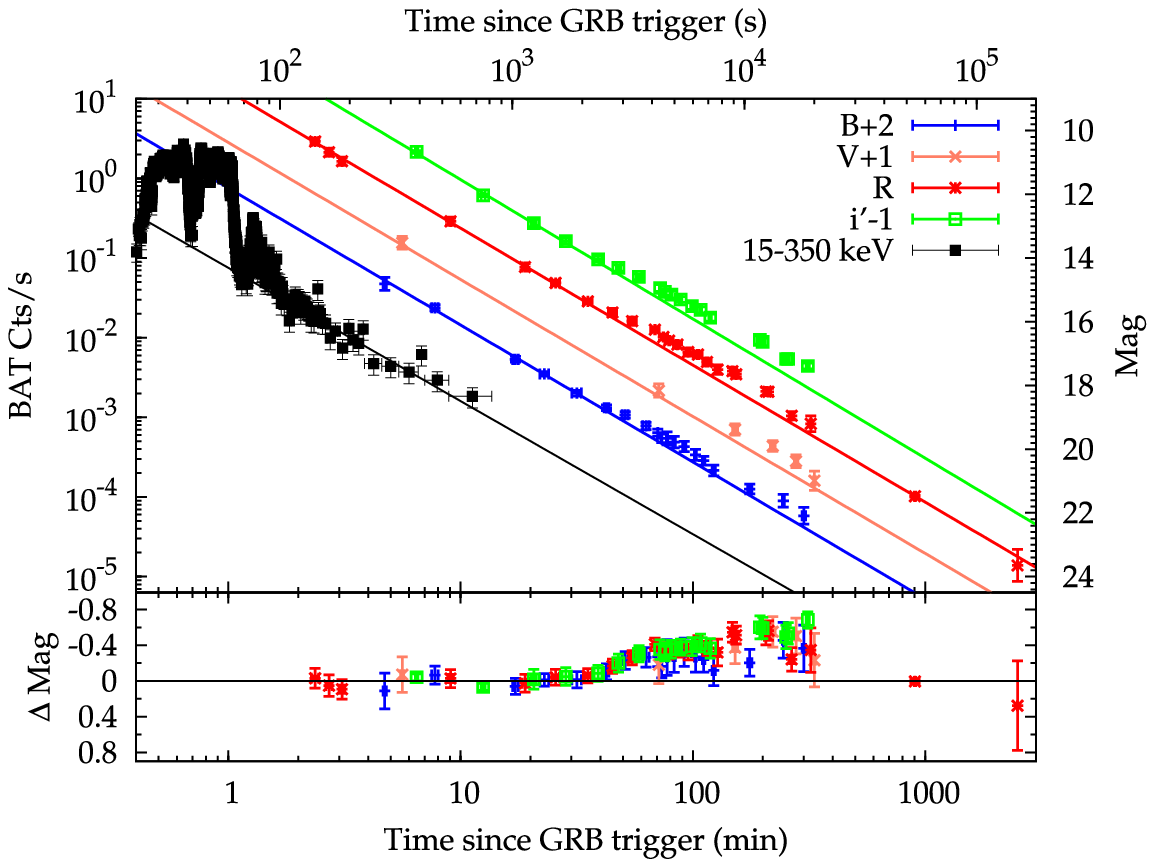}{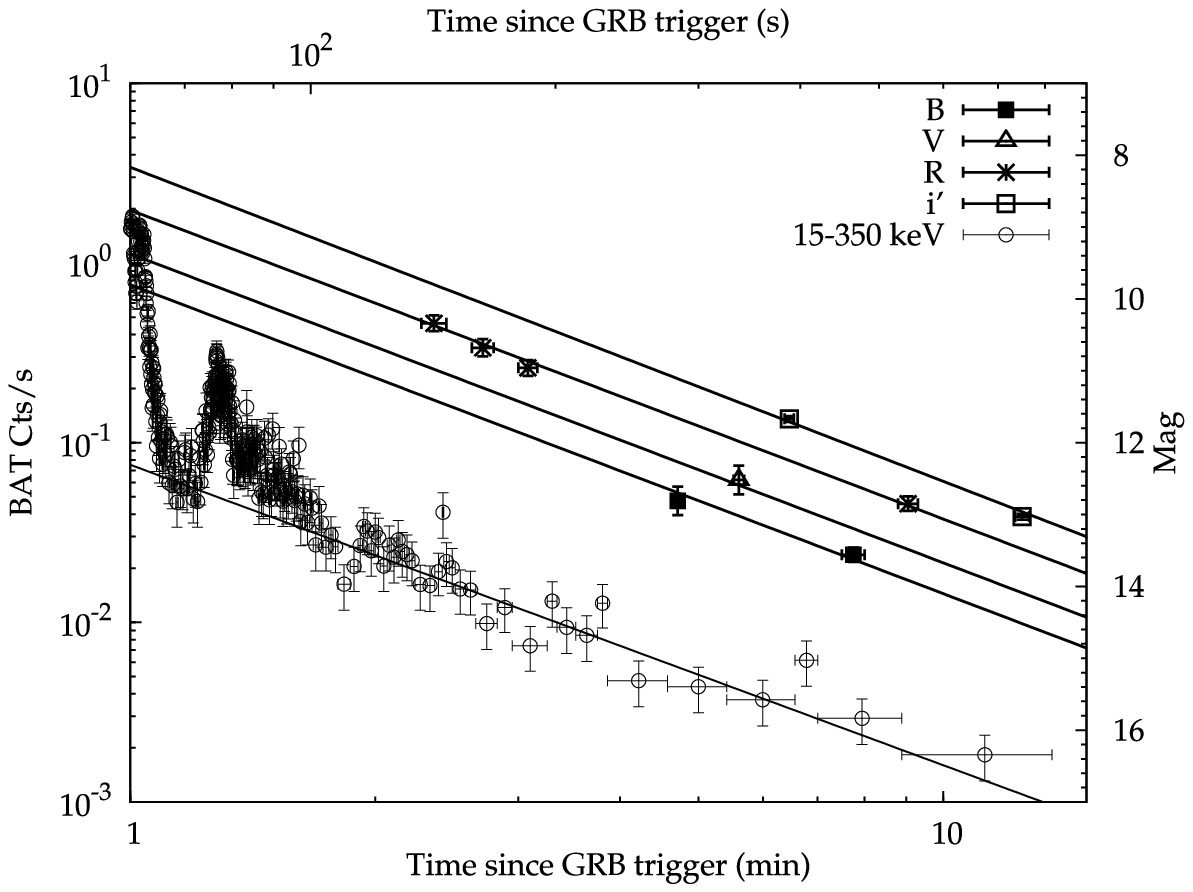} 
\caption{{\bf Left:}
FTS BVR$i^\prime$ optical and {\em Swift} $\gamma$-ray light curves of 
GRB~061007. Optical light curves are fitted with a single power-law fits for t$<40$ min and the background-subtracted $\gamma$-ray light curve is fitted with a similar power law to data at $t>120$~s. The BAT count
rate (counts per sec per fully illuminated detector for an equivalent on-axis
source) and optical magnitudes are given on the left and
right vertical axes respectively. The lower panel
shows the optical residuals after removal of the corresponding power law (see Table~\ref{tab:alphas}), highlighting the
presence of a broad bump in all filters for $t>40$~min. Late time
R-band photometry from VLT and Magellan images are included at t$\sim$1000
 and t$\sim$2500 min respectively and are consistent with extrapolation of the power-law decay.  
 {\bf Right:} An expanded
view of the optical and $\gamma$-ray light curves at early time (1$<$t$<$11 min).
 \label{fig:LC}}
\end{figure}

\begin{figure}
\plotone{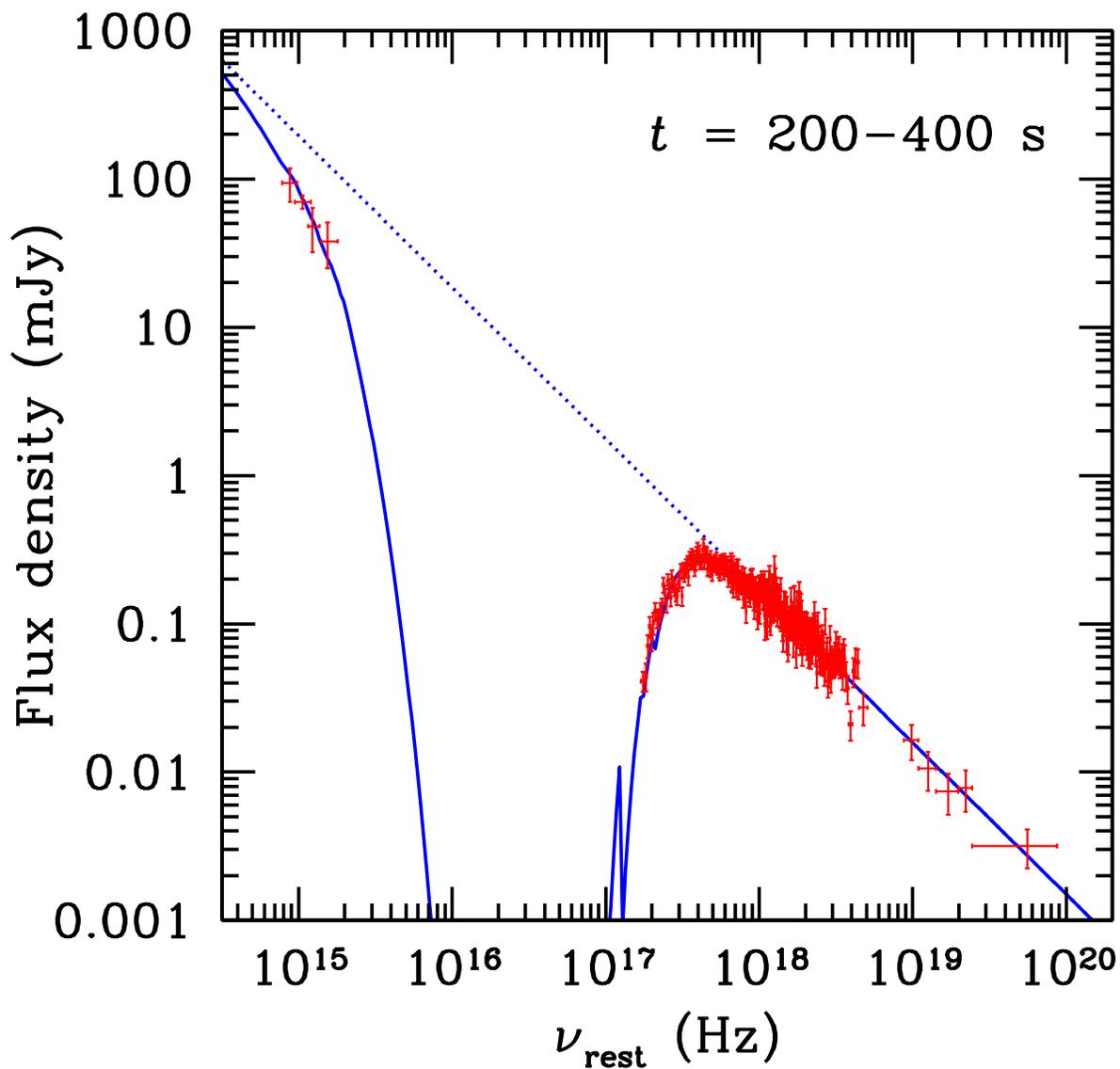}
\caption{Broad-band optical to $\gamma$-ray spectral energy
distribution derived for the time interval $200<t_{obs}<400$ s,
fitted with an absorbed power law with $\beta$(opt-X-$\gamma$)~=~1.02$\pm$0.05 and rest frame extinction $A_V$(SMC)=0.48$\pm$0.19 mag (solid line). The unabsorbed power law is also shown (dotted line).\label{fig:SED}}
\end{figure}

Figure~\ref{fig:LC} shows the FTS BVR$i^\prime$ and
background-subtracted {\em Swift}-BAT light curves of GRB~061007; a
single power-law decay with $\alpha=1.72\pm0.10$
(F~$\propto$~t$^{-\alpha}$) provides a good fit to all optical light
curves for $t<40$~min, resulting in small residuals in this time range
and emphasizing the presence of an additional broad, multicolor bump
over the time range $40<t<350$ min (see
Table~\ref{tab:alphas}). Late-time VLT R-band photometry confirms that
the underlying power-law decay continues, showing no evidence for a
break before 1000 min (0.624 d). This is consistent with the
continuation of the simple power-law decay in the XRT light
curve~\citep{Schady06c} .  Similarly, the marginal Magellan detection
extends the continuation of the unbroken power-law decay to $t=2495$
min (1.73 d), and, combined with the final upper limit R$>23.3$ mag at
4087 min (2.84 d - Table \ref{tab:obsfts}), constrains the brightness
of the host galaxy. A similar power law ($\alpha=1.7\pm0.1$) also
provides a good fit to the decay of the long, soft tail of the
 $\gamma$-ray emission for $t>120$~s, although slight modulation of
 the power law due to the intrinsic variability is evident, as
 illustrated in the expanded view of the light curves of the first 10
 min (Figure~\ref{fig:LC} right panel; Table~\ref{tab:alphas}). A
 similar deviation from the R-band power law may be present at
 $\sim$2.3 min (see residuals Figure~\ref{fig:LC}), coincident in time
 with a small $\gamma$-ray flare, but the effect is marginal.

The broad-band spectral energy distribution shown in Figure
\ref{fig:SED} confirms the overall common power-law decay in all bands
from $\gamma$ to optical and is well fitted by an absorbed power law
with optical extinction $A_V\sim0.48\pm0.19$ mag with an SMC
profile. Comparing this with the $N_{\rm H}$ derived from the X-ray
spectrum, we find a deficit of optical extinction, as already found in
most GRBs (Stratta et al. 2004; Kann et al. 2006; Galama \& Wijers
2001).  In principle this could be explained in a number of ways (see
e.g. the case of GRB~051111, Guidorzi et al. 2006b): 1) an extinction
profile different from those typical of MW, SMC, LMC (e.g. Stratta et
al. 2004; Savaglio \& Fall 2004); 2) a population
of dust grains skewed towards big sizes, possibly as a result of dust
destruction due to the GRB itself (e.g. Maiolino et al. 2001; Chen et
al. 2006); 3) a significant presence of
molecular gas \citep{Arabadjis99}; 4) overabundance of some alpha
metals responsible for absorption in X-rays (e.g., GRB~050401, Watson
et al. 2006).

Below, we explain the observed multiwavelength evolution of GRB~061007
at $t>70$ s with an expanding fireball whose optical, X-ray and
$\gamma$-ray emission is dominated by synchrotron emission from a
forward shock whose cooling frequency lies above the X-ray band and
typical synchrotron frequency lies just below the optical band; in
contrast, the typical frequency of the reverse shock lies in the radio
band at early time, with the typical frequency of the forward shock
entering the radio band at $\sim$2.5~days after the burst.

\subsection{Forward Shock Emission from a Decelerating Fireball}

The optical and X-ray afterglow light curves are described by the same
power law $\alpha \sim 1.7$ to more than $10^5$ sec after the burst;
this indicates that these photons are radiated from the same forward
shock, and optical and X-ray bands are in the same domain of a
synchrotron spectrum\footnote{Even if we assume reverse shock
emission, the observed bands should be in the same spectral domain
$\nu_m < \nu_{opt}, \nu_X < \nu_c$, because reverse shock emission
above the cooling frequency vanishes after the shock crossing. In this
case we obtain the theoretical value $\alpha-3\beta/2 \sim 1$ which is
not consistent with observation $\alpha-3\beta/2\sim 0$}.  In
principle a bump of a forward shock (blast wave) light curve, as seen
for t$>40$ min (Figure~\ref{fig:LC}), can be produced by inhomogeneity
of the ambient medium or by energy injection into a blast
wave. However, energy injection causes a transition from a blast wave
to another with a larger energy, and it leads to a shift of the
afterglow decay baseline after the bump (at this late time the
evolution of a blast wave is adiabatic). Since the optical flux comes
back to the pre-bump decay line, inhomogeneity of ambient medium is
favored to explain the bump around 100 min in the optical. The X-ray
band (and the optical band) should be below the synchrotron cooling
frequency $\nu_c$, otherwise the inhomogeneity in the ambient medium
does not produce an afterglow bump (Kumar 2000). For the synchrotron
shock model, the decay index and spectral index at frequencies $\nu_m
< \nu < \nu_c$ are $\alpha=3(p-1)/4$ and $\beta=(p-1)/2$, respectively
(Sari, Piran and Narayan 1998). With $p\sim 3$ these give reasonable
fits to the observed indices. We note that if a burst occurred in a wind
medium with density $n \propto R^{-s}$, the decay index in this spectral domain
$\nu_m < \nu <\nu_c$ is given by $\alpha=[2s+3(p-1)(4-s)]/(16-4s) \sim
1.7$ for $p=3$ and $s=1$ (Monfardini et al. 2006). The additional
parameter $s$ allows a better fits to the decay index, but still a
rather large value of $p\sim3$ is required to explain the observed
$\beta$.

Since the optical and X-ray light curves show no break during the
observations, the typical synchrotron frequency $\nu_m \propto t^{-3/2}$
should be already below the optical band $\nu_m<\nu_{opt} \sim 5\times
10^{14}$ Hz  before our earliest optical observation was made at 
$t=137$ sec, while the cooling frequency $\nu_c \propto t^{-1/2}$ should
be still above X-ray band $\nu_X < \nu_c \sim 10^{18}$ Hz at $t \sim
10^5$ sec. Using the results in Sari, Piran and Narayan (1998), we get
\begin{eqnarray}
\nu_{m}(t) &\sim& 1.0 \times 10^{21} \zeta^{1/2} \epsilon_B^{1/2}
\epsilon_e^2 E_{54}^{1/2} t_m^{-3/2} ~\mbox{Hz},\\
\nu_{c}(t) &\sim& 6.8\times10^{12}  \zeta^{-1/2}
\epsilon_B^{-3/2} E_{54}^{-1/2} n^{-1} t_m^{-1/2}
~\mbox{Hz}
\end{eqnarray}
where $\zeta=(1+z)/2.26$, $E_{54}=E/10^{54}$ ergs, 
$t_m$ is the observer's time in units of min, $n$ is the density in cm$^{-3}$, $\nu_m$ is  
proportional to $(p-2)/(p-1)$ and we have assumed $p=3$.
The requirements for the break
frequencies at $t=137$ s and at $t=10^5$ s give constraints on the
microphysics parameters, 
\begin{eqnarray}
\epsilon_B &<& 3.0 \times 10^{-5} n^{-2/3}
\paren{\zeta E_{54}}^{-1/3} \\
\epsilon_e &<& 1.7\times10^{-2}  \paren{\frac{\epsilon_B}{3.0\times10^{-5}}}^{-1/4}
\paren{\zeta E_{54}}^{-1/4} 
\end{eqnarray}

These energy partition values are somewhat small but are not
inconsistent with values derived for other bursts (e.g. Panaitescu \&
Kumar 2002).\\

\subsubsubsection{Onset of the Afterglow and Lack of Optical Flash}

ROTSE detected a 13.6-mag optical counterpart at 26 s after the GRB
trigger (Rykoff et al. 2006). This magnitude is dimmer than the
extrapolation of our observations. It indicates that the afterglow
peaked between 26 s and 137 s, and very likely around the end of the
main pulses of the prompt emission $t_p \sim 100$ s. Using this peak
time, we can estimate the Lorentz factor of the fireball at that time
(Sari and Piran 1999). It is \begin{equation} \Gamma \sim
\paren{\frac{3(1+z)^3E}{32\pi n m_pc^5 t_p^3}}^{1/8} \sim 230
\paren{\frac{t_p}{10^2 sec}}^{-3/8} \paren{\zeta^3 E_{54}/n}^{-1/8}
\end{equation} It is interesting that the tail of the $\gamma$-ray
emission is described by the same power-law decay $\alpha\sim1.7$ as
the optical and X-ray emission (Figure~\ref{fig:LC}); indeed the flux
at the dip in the prompt phase at $\sim$40 s is consistent with the
underlying power-law line. If the afterglow started at $\sim 30$ s,
the Lorentz factor is $\sim 360$. No jet break within $10^5$ s
\footnote{Another possible scenario is that a jet break happened
immediately after the prompt phase at $t<100$ sec. This scenario
requires a very small electron index to explain the decay indices of
the optical and X-ray light curves.}  gives a lower limit of the GRB
jet opening angle $\theta_0(rad) > 0.07 ~n^{1/8}
E_{54}^{-1/8}\zeta^{-3/8} (t/10^5 sec)^{3/8}$.

There is no reverse shock emission component even though the optical
observations started right after the prompt emission phase.  We can
explain the lack of optical flash naturally in the standard model if
the typical frequency of the forward shock emission is lower than
optical band $\nu_m < \nu_{opt}$ at the onset of afterglow (the peak
time $t_p$).  At the peak time (the shock crossing time), the forward
and reverse shocked regions have the same Lorentz factor and internal
energy density, the cooling frequency of the reverse shock is equal to
that of the forward shock $\nu_{c,r}(t_p) \sim \nu_{c}(t_p)$. The
matter density in the reverse shocked region is much higher than in
the forward shock region, and it makes the electron temperature
lower. The typical frequency of the reverse shock is much lower than
that of the forward shock $\nu_{m,r}(t_p) \ll \nu_m(t_p) < \nu_{opt}$
(e.g. Kobayashi \& Zhang 2003).

The shock regions have the same amount of shock energies at $t_p$, and 
 their luminosities ($\sim \nu F_\nu$ at $\nu_c$) are comparable at
 $t_p$. Considering that the two emission components have the same
 spectral index, at $t_p$ they contribute equally to the flux at any
 observed frequency between $\nu_{m,f}$ and $\nu_c$.  After the shock
 crossing $t>t_p$ the reverse shock emission decays more rapidly as
 $\sim t^{-2}$, so the contribution from the reverse shock should be
 masked by the forward shock component in the decay phase.

Forward shock emission peaks when a fireball decelerates or when the
typical frequency passes through the observed band. If an optical peak is
not associated with a spectral change from $\nu^{1/3}$ to
$\nu^{-(p-1)/2}$, it could be because of the deceleration of a
fireball, and the typical frequency could be located below the
observed band at that time. In such a case, the optical flash should
always be absent. In recent years, ground-based robotic telescopes
reported the lack of optical flashes, except for a few cases.  The
lack might be explained with low $\nu_m$ originating from rather small
microphysics parameters, $\epsilon_e$ and $\epsilon_B$. GRB 060117 and
GRB 061007 might be clear example cases.

\subsection{Implications for Late-Time Radio Emission}

After the deceleration, the emission frequency of each electron in the
shocked ejecta (reverse shock region) drops quickly with time as
$\nu_{e} \propto t^{-73/48}$.  Both the typical frequency and cooling
frequency drop with this scaling, because after a reverse shock has crossed the ejecta (deceleration), no new electrons are injected and all electrons cool by
adiabatic expansion only. The peak power decays as $F_{\nu,max} \propto
t^{-47/48}$. The flux below or above the typical frequency $\nu_{m,r}$
evolves as (Sari \& Piran 1999; Kobayashi \& Sari 2000, their Eq. 3),
\begin{equation} \label{emission} F_{\nu} \propto \left\{
                \begin{array}{@{\,}ll}
  t^{-17/36} & \nu < \nu_{m,r},  \\  
  t^{-(73p+21)/96}\propto t^{-5/2}  ~(p=3)     & \nu > \nu_{m,r}.
                \end{array}
                \right. 
\end{equation}
The optical afterglow of GRB 061007 was very bright. If the forward
shock typical frequency is close to the optical band at the deceleration
time $\nu_m(t_p) \sim 5\times 10^{14}$ Hz, extrapolating our
observational results to the deceleration time $t_p \sim 100$
s, we obtain the peak flux of  the forward shock emission $F_{\nu,max}
\sim 420$ mJy. 
The typical frequency of the reverse shock emission is lower
by a factor of $\sim \Gamma^2$ (Kobayashi \& Zhang 2002), and 
 it is in the radio band $\nu_{m,r}(t_p)\sim \nu_m/\Gamma^2 \sim 9.5
(\Gamma/230)^{-2}$ GHz. At the deceleration time, the peak flux of the
reverse shock emission is larger by  a factor of $\Gamma \sim 230$ 
than that of the forward shock emission. 
 Since the reverse shock emission in the radio band  initially 
decays as $t^{-17/36}$ until the typical frequency passage, then as
$t^{-5/2}$, the reverse shock radio emission at 1 day is 
about $\sim 8.6 ~(\nu/4.8\mbox{GHz})^{-1}(t/1\mbox{day})^{-5/2} \mu $Jy. 

The typical frequency of the forward shock emission $\nu_m \propto
t^{-3/2}$ comes to radio band $\nu =4.8$ GHz  
 about $2.5$ day after the burst.
Before the passage, the flux increases as $t^{1/2}$, and then 
decays as $t^{-1.7}$. The peak value is expected  to be $\sim 420$ mJy. 
At low frequencies and early times, self absorption plays an important
role and significantly reduces the afterglow flux. A simple 
rough estimate of the maximum flux is the emission from a black
body with the shock temperature (e.g. Sari \& Piran 1999; Kobayashi \&
Sari 2000).  The temperature is given by the random energy of the typical
electron $k_B T \sim m_ec^2\gamma_{m} \sim \epsilon_e m_pc^2 \Gamma/2$ for p=3.
If the observed frequency is above the typical frequency, the electrons 
radiating into the observed frequency have an energy higher than the typical energy, $m_ec^2\gamma_m$, by a factor of
$(\nu/\nu_{m})^{1/2}$. Following the notation of Sari \& Piran (1999), the upper limit of blackbody emission is
\begin{eqnarray}
F_{\nu}^{BB}= \pi (R_\perp/d)^2 \Gamma S_{\nu}
\end{eqnarray}
where $d=d_L/(1+z)^{1/2}$ and $S_\nu=(2\nu^2/c^2)k_B T$. 
Therefore
\begin{eqnarray}
F_{\nu}^{BB} &\sim&  \pi (1+z) m_p  \nu^2 \epsilon_e \Gamma^2 
\paren{\frac{R_\perp}{d_L}}^2, \\
&\sim& 6 ~ 
\paren{ \frac{\nu}{\mbox{4.8GHz}}}^{2} \paren{\frac{t}{\mbox{1day}}}^{1/2}
\epsilon_{e,-2} \paren{E_{54}/n}^{1/2}
~\mbox{mJy} 
\end{eqnarray}
where $\epsilon_{e,-2}=\epsilon_e/10^{-2}$ and $R_\perp \sim 4.6\Gamma c t$ is the observed size of the fireball, assuming
$z=1.26$ and $d_L=2.7\times10^{28}$ cm.
Note that other emission estimates use only scalings (normalizations
are given by the observations), the blackbody upper limit is more model
dependent.

The results from ATCA (van der Horst et al. 2006a,b) give even lower
limits about $50\mu$Jy level. Their formal flux measurements are $-25
\pm 45 \mu$ Jy (1.00-1.24 days after the burst) and $-1 \pm 43\mu$Jy
(5.03-5.28 days after) at 4.8 GHz. Although the blackbody limit is
possibly less reliable, the radio non-detection might imply a
high-density environment $n\gg 1$. The peak flux of the forward shock
emission is given by (Sari, Piran \& Narayan 1998), \begin{equation}
F_{\nu,max}(z=1.26) \sim
20~\paren{\frac{\epsilon_B}{5\times10^{-5}}}^{1/2}
E_{54}n^{1/2}~\mbox{mJy} \end{equation} 
As can be seen from Equations 3 and 10, the dependence of $F_{\nu,max}$ on density is weak ($\sim n^{1/6}$), and thus the bright afterglow cannot be easily explained by a high density environment.

\subsection{GRB\,061007 and Cosmological Spectral--Energy Correlations}

{\em Swift} GRB~061007 is one of the most luminous GRBs ever detected
and thus, with its high-quality $\gamma$-ray data, well-characterised
light curves and well-determined peak energy, it is an ideal object to
test the cosmological spectral--energy correlations, namely the
$E_{\rm peak}$-$E_{\rm iso}$ (Amati et al. 2002), $E_{\rm
peak}$-$E_{\gamma}$ (Ghirlanda et al. 2004) and $E_{\rm peak}$-$E_{\rm
iso}$-$t_{\rm b}$ (Liang \& Zhang 2005) correlations.  As described in
Section~\ref{sec:obs} the prompt $\gamma$-ray emission at Mev energies
was detected by three additional satellites which provide estimates of
the peak energy.  While the peak energies measured by Konus/{\em Wind}
and {\em RHESSI} are comparable, that measured by {\em Suzaku}/WAM is
significantly larger; in addition to possible systematics, this may be
explained by the fact that this instrument measured the brightest part
of the GRB, as suggested by their low duration estimate (T90$\sim$50~s
vs 90~s measured by Konus/{\em Wind})\footnote{The lower bound of the
energy band of WAM is somewhat higher than that of Konus/{\em Wind}
and {\em RHESSI}, which may result in an overestimate of $E_{\rm
peak}$ (Amati~2006).}. We therefore use the spectral information from
Konus/{\em Wind} and {\em RHESSI}, conservatively assuming an $E_{\rm
peak}$ range including both 90\% c.l. intervals provided by the two
instruments, i.e. $E_{\rm peak}$=394$\pm$55~keV, corresponding to a
cosmological rest-frame peak energy of 890$\pm$124 keV.

The isotropic--equivalent radiated energy is $E_{\rm
iso}$=(1.0$\pm$0.1)$\times$10$^{54}$ erg in the 1-10000 keV
cosmological rest--frame, assuming H$_0$ = 70 km/s/Mpc,
$\Omega_{\Lambda}$ = 0.3 and $\Omega_m$ = 0.7 and a Band spectral
shape \citep{Band93} with parameters and fluence provided by
Konus/{\em Wind} (Golenetskii at al. 2006). Following Amati
(2006) the $E_{\rm peak}$-$E_{\rm iso}$ correlation predicts a
corresponding $E_{\rm peak}$ = 907~keV, fully consistent with the
measured value for GRB~061007. Thus, GRB~061007 further confirms the
validity of the $E_{\rm peak}$-$E_{\rm iso}$ correlation in the very
high radiated energy regime and since this correlation should only
hold for prompt emission, confirms that GRB~061007 has prompt emission
typical of other {\em Swift} bursts on the correlation, despite its
unusually simple afterglow emission.

Finally, the test of the $E_{\rm peak}$-$E_{\gamma}$ and $E_{\rm
peak}$-$E_{\rm iso}$-$t_{\rm b}$ correlations, which is of particular
importance given their possible use for the estimate of cosmological
parameters (e.g. Firmani et al. 2005) requires the detection of a late
break in the afterglow light curves. As can be seen in Figure 1, our
optical light curves show no evidence of a break up to $\sim150$~ks
and the XRT light curve continues unbroken to at least $10^6$~s
(Schady et al. 2006c), providing a firm lower limit to the jet break
time. Combining our estimates of $E_{\rm iso}$ and the lower limit to
the jet opening angle of $\sim 4.7^{\circ}$ inferred from the optical
light curve, we derive a collimation-corrected energy $E_{\gamma}>
3.4\times$10$^{51}$ erg; this limit would predict $E_{\rm peak}>$~1274
keV in order to be consistent with the $E_{\rm peak}$-$E_{\gamma}$
correlation \citep{Nava06}, a value somewhat higher than the measured
value of 890$\pm$124 (90\% c.l.). By accounting for the logarithmic
dispersion of this correlation \citep{Nava06}, the deviation of the
measured value from this lower limit is $\sim$1.6~$\sigma$. Better
consistency is found with the $E_{\rm peak}$-$E_{\rm iso}$-$t_{\rm b}$
correlation, which, when adopting the parameters reported by Liang \&
Zhang (2005) predicts a lower limit to $E_{\rm peak}$ of 986
keV. Using the lower limit to the jet break time $t\sim10^6$ s
suggested by the X-ray light curve, we conclude that GRB~061007
becomes a secure outlier to both E$_{\rm peak}$-${\rm E}_\gamma$ and
E$_{\rm peak}$-${\rm E}_{\rm iso}$-$t_b$ correlations, deviating by more than $3~\sigma$. X-ray light curves for other
 GRBs suggest the existence of a population of objects similar to
 GRB~061007 that are inconsistent with these correlations (Sato et
 al. 2006; Willingdale et al. in prep.), although optical observations
 were not available for these objects. GRB~061007 might therefore
 transpire to be the norm rather than an anomaly.

\section{Conclusions} \label{sec:conc}

We have presented a multiwavelength study of the very bright
GRB~061007 based on optical observations that cover the period from
137 s to 3 days after the burst and {\em Swift} XRT and BAT data.  We
conclude that the afterglow commences as early as $30-100$ s after the
onset of the GRB and the optical, X-ray and underlying $\gamma$-ray
emission, which are described by the same power-law decay $\alpha \sim
1.7$ after this time, represent photons that are radiated from the
same forward shock, with optical and X-ray bands in the same domain of
a synchrotron spectrum. In contrast, the typical frequency of the
reverse shock emission is already in the radio domain at early time,
explaining the lack of a bright early-time optical flash and late time
radio flare.  We note the
similarity of the light curves of GRB~061007 to that of GRB~060117,
which was only observed in a single optical filter~\citep{Jelinek06}
and suggest it too may represent the onset of an early afterglow
similar to GRB~061007.  Finally, we highlight the simplicity of the
early-onset afterglow of GRB~061007, which with its well-behaved light
curves, typical circumburst environment and well-determined peak
$\gamma$-ray energy, make it a useful candidate for inclusion in
current cosmological correlations. We find that this event is fully
consistent within 1 $\sigma$ with the E$_{\rm peak}$-${\rm E}_{\rm
iso}$ correlation and also potentially consistent within $1.6~\sigma$
with the E$_{\rm peak}$-${\rm E}_\gamma$ and E$_{\rm peak}$-${\rm
E}_{\rm iso}$-$t_b$ correlations using a lower limit to the
jet break time $t_b$, and thus to E$_\gamma$,  derived from
our data. We note that since the X-ray afterglow continued without a jet
break until $t>10^6$ s, this event is firm outlier to
the E$_{\rm peak}$-${\rm E}_\gamma$ and E$_{\rm peak}$-${\rm E}_{\rm
iso}$-$t_b$ correlations, deviating from these correlations by more than $\sim3~\sigma$.

\acknowledgements CGM acknowledges financial support from the Royal
Society. AXM, MFB, ER, PTO, acknowledge funding from the Particle
Physics and Astronomy Research Council (PPARC). RoboNet1.0 and Swift
are supported by PPARC.  The Faulkes Telescopes, now owned by Las
Cumbres Observatory, are operated with support from the Dill Faulkes
Educational Trust. This work is partly based on data taken with ESO
telescopes at VLT under program 078.D-0752. We thank the kind
assistance of the observing staff at Paranal. We thank J.E. Rhoads for
assistance in acquiring the Magellan data, L. Nava for input on the
spectral correlations and the anonymous referee for useful comments
and suggestions.

\clearpage


\begin{deluxetable}{|c|c|c|c|c|}
\tablewidth{0pt}
\tablecaption{FTS, VLT and Magellan optical observations of GRB~061007.\label{tab:obsfts}}
\tablehead{
\colhead{\textsc{Telescope}} & 
\colhead{\textsc{$\Delta$t}} & 
\colhead{\textsc{Filter}} & 
\colhead{\textsc{Exposure}} & 
\colhead{\textsc{Mag $\pm$ Err\tablenotemark{\dagger}}}\\
\colhead{\textsc{}} & 
\colhead{\textsc{(min)}} & 
\colhead{\textsc{}} & 
\colhead{\textsc{Time (s)}} & 
\colhead{\textsc{}}
}
\startdata
FTS &     2.28  & R$_{\rm C}$  &   10.0 &  10.34 $\pm$ 0.11\\
 &     2.63	 & R$_{\rm C}$  &	  10.0 &  10.68 $\pm$ 0.12\\
 &     3.00	 & R$_{\rm C}$  &	  10.0 &  10.96 $\pm$ 0.11\\
 &    4.63 	 & B  &	  10.0 &  12.81 $\pm$ 0.20\\
 &     5.52	 & V  &	  10.0 &  12.52 $\pm$ 0.20\\
 &     6.38	 &    SDSS-$i^\prime$  &	  10.0 &  11.67 $\pm$ 0.04\\
 &     7.50	 & B  &	  30.0 &  13.56 $\pm$ 0.10\\
 &     8.80	 & R$_{\rm C}$  &	  30.0 &  12.85 $\pm$ 0.10\\
 &    12.25	 &    SDSS-$i^\prime$  &	  30.0 &  13.03 $\pm$ 0.04\\
 &    16.73	 & B  &	  60.0 &  15.18 $\pm$ 0.09\\
 &    18.42	 & R$_{\rm C}$  &	  60.0 &  14.28 $\pm$ 0.10\\
 &    20.18	 &    SDSS-$i^\prime$  &	  60.0 &  13.90 $\pm$ 0.11\\
 &    21.93	 & B  &	 120.0 &  15.64 $\pm$  0.07\\
 &    24.63	 & R$_{\rm C}$  &	 120.0 &  14.78 $\pm$  0.08\\
 &    27.38	 &    SDSS-$i^\prime$  &	 120.0 &  14.47 $\pm$  0.10\\
 &    30.15	 & B  &	 180.0 &  16.24 $\pm$  0.09\\
 &    33.78	 & R$_{\rm C}$  &	 180.0 &  15.36 $\pm$  0.08\\
 &    37.55	 &    SDSS-$i^\prime$  &	 180.0 &  15.04 $\pm$  0.09\\
 &    41.45	 & B  &	 120.0 &  16.70 $\pm$  0.09\\
 &    44.08	 & R$_{\rm C}$  &	 120.0 &  15.71 $\pm$  0.08\\
 &    46.83	 &    SDSS-$i^\prime$  &	 120.0 &  15.31 $\pm$  0.10\\
 &   49.63 	 & B  &	 180.0 &  16.92 $\pm$  0.10\\
 &    53.28	 & R$_{\rm C}$  &	 180.0 &  15.98 $\pm$  0.07\\
 &    57.07	 &    SDSS-$i^\prime$  &	 180.0 &  15.59 $\pm$  0.09\\
 &    60.85	 & B  &	 240.0 &  17.27 $\pm$  0.11\\
 &    68.40	 & R$_{\rm C}$  &	  30.0 &  16.25 $\pm$  0.09\\
 &    70.50	 & B  &	  10.0 &  17.48 $\pm$  0.11\\
 &    71.23	 & V  &	  10.0 &  17.16	$\pm$  0.21\\
 &    72.13	 &    SDSS-$i^\prime$  &	  10.0 &  15.94 $\pm$  0.11\\
 &    73.17	 & B  &	  30.0 &  17.66 $\pm$  0.13\\
 &    74.40	 & R$_{\rm C}$  &	  30.0 &  16.48 $\pm$  0.09\\
 &  75.62  	 &    SDSS-$i^\prime$  &	  30.0 &  16.06 $\pm$  0.10\\
 &    76.90	 & B  &	  60.0 &  17.66 $\pm$  0.20\\
 &    78.53	 & R$_{\rm C}$  &	  60.0 &  16.59 $\pm$  0.09\\
 &    80.32	 &    SDSS-$i^\prime$  &	  60.0 &  16.15 $\pm$  0.10\\
 &   82.08 	 & B  &	 120.0 &  17.78 $\pm$  0.18\\
 &   84.77 	 & R$_{\rm C}$  &	 120.0 &  16.72 $\pm$  0.09\\
 &  87.52  	 &    SDSS-$i^\prime$  &	 120.0 &  16.30 $\pm$  0.09\\
 &   90.30 	 & B  &	 180.0 &  17.92 $\pm$  0.16\\
 &    93.95	 & R$_{\rm C}$  &	 180.0 &  16.95 $\pm$  0.09\\
 &    97.73	 &    SDSS-$i^\prime$  &	 180.0 &  16.51 $\pm$  0.10\\
 &   101.55	 & B  &	 120.0 &  18.18 $\pm$  0.17\\
 &  104.20 	 & R$_{\rm C}$  &	 120.0 &  17.03 $\pm$  0.08\\
 &  106.93 	 &    SDSS-$i^\prime$  &	 120.0 &  16.63 $\pm$  0.10\\
 &   109.68	 & B  &	 180.0 &  18.36 $\pm$  0.14\\
 &   113.43	 & R$_{\rm C}$  &	 180.0 &  17.26 $\pm$  0.08\\
 &   117.20	 &    SDSS-$i^\prime$  &	 180.0 &  16.87 $\pm$  0.11\\
 &   120.98	 & B  &	 240.0 &  18.67 $\pm$  0.17\\
 &   125.62	 & R$_{\rm C}$  &	 240.0 &  17.50 $\pm$  0.15\\
 &	 145.55	 & R$_{\rm C}$  &  300.0 &  17.55 $\pm$ 0.11\\
 &	 150.92	 & R$_{\rm C}$   & 300.0 &  17.65 $\pm$ 0.10\\
 &	 160.12	 & V   & 450.0 &  18.38 $\pm$ 0.18\\
 &	171.98 	 & B   & 450.0 &  19.25 $\pm$ 0.15\\
 &    191.77 &     SDSS-$i^\prime$  & 300.0 &  17.57 $\pm$ 0.13\\
 &	  197.15 &     SDSS-$i^\prime$   & 300.0 & 17.63 $\pm$ 0.12\\
 &	  203.75 &  R$_{\rm C}$   & 300.0 & 18.19 $\pm$ 0.11\\
 &	  209.13 &  R$_{\rm C}$   & 150.0 & 18.20 $\pm$ 0.11\\
 &	  215.67 &  V   & 600.0 & 18.90	$\pm$ 0.17\\
 &	  242.75 &  B   & 300.0 & 19.62 $\pm$ 0.20\\
 &	  249.33 &     SDSS-$i^\prime$   & 300.0 & 18.17 $\pm$ 0.13\\
 &	  254.72 &     SDSS-$i^\prime$   & 300.0 & 18.17 $\pm$ 0.12\\
 &	  261.28 &  R$_{\rm C}$   & 600.0 & 18.95 $\pm$ 0.13\\
 &	  273.20 &  V   & 600.0 & 19.38	$\pm$ 0.20\\
 &	  295.43 &  B   & 600.0 & 20.09 $\pm$ 0.26\\
 &	  307.42 &     SDSS-$i^\prime$   & 600.0 & 18.39 $\pm$ 0.15\\
 &    319.37 &  R$_{\rm C}$   & 300.0 & 19.20 $\pm$ 0.25\\
 &	  331.28 &  V   & 300.0 & 19.99	$\pm$ 0.30\\
VLT &    899.58 &  R$_{\rm C}$   & 600.0 & 21.48 $\pm$ 0.03\\
Magellan &    2494.80&  R$_{\rm C}$   & 240.0 & 23.65 $\pm$ 0.50\\
 &    4086.64&  R$_C$   & 240.0 & $>$23.30\\
\enddata\\
\tablenotetext{\dagger}{Quoted errors for BV$i^\prime$ filters are statistical. Absolute calibration requires inclusion of the systematic error 0.25~mag due to the uncertainty in the USNO B1 photometric calibration. This error has been fully included when deriving the SED (See Figure~\ref{fig:SED} and Table \ref{tab:sedfits}).}
\end{deluxetable}

\clearpage

\begin{table}
\begin{center}
\caption{Results of power-law fits to optical and $\gamma$-ray light curves as shown in Figure~\ref{fig:LC}.\label{tab:alphas}}
\begin{tabular}{ccc}
  \hline
Band	& Time Range For Fit & Temporal Power-Law Decay ($\alpha$)\\
\hline
B	& $t<40$~min    & 1.72$\pm$0.05\\
V	&  $-$	     &	Fixed to $\alpha_{\rm R}$\\
R	& $t<40$~min &  1.72$\pm$0.01\\
$i^\prime$	& $t<40$~min   &  1.75$\pm$0.06\\
BAT($15-350$~keV)& $t>40$~s & 1.7$\pm$0.1\\
  \hline
 \end{tabular}								
 \medskip
\end{center}							
\end{table}

\begin{table}
\begin{center}
\caption[]{Best-fitting parameters derived from optical-X-$\gamma$ spectral energy
distribution.\label{tab:sedfits}} 
\begin{tabular}{lcccc}
  \hline
		& SMC\tablenotemark{\ddag} & LMC & MW & SMC, $\beta_{X\gamma}$ fixed\\
\hline
$\beta$		& 1.02$\pm$0.05 & 1.03$\pm$0.05 & 1.03$\pm$0.05& 1.01\\
$A_V$ (mag)	& 0.48$\pm$0.19  & 0.54$\pm$0.20  & 0.54$\pm$0.20& 0.29$\pm$0.06\\
$\chi^2$/d.o.f	& 195/174	& 195/174       &195/174& 197/175\\
  \hline
 \end{tabular}	
\tablenotetext{\ddag}{See Figure~\ref{fig:SED}}
\end{center}		
\end{table}

\clearpage


\begin{figure}
\epsscale{1.0}
\plotone{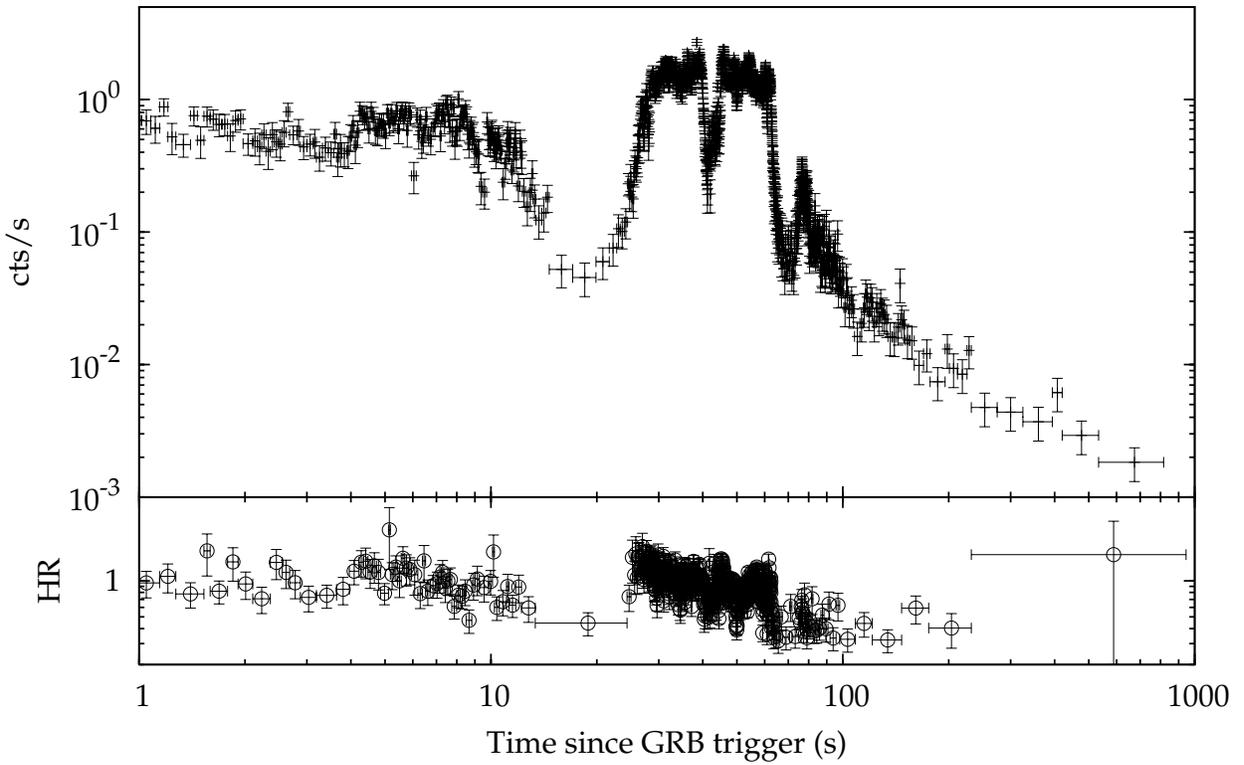}
\caption{Evolution of the hardness ratio (HR), derived from comparison of count rates in 50$-$350 and 15$-$50~keV bands, compared with the BAT (15$-$350~keV), showing a marginal softening of the $\gamma$-ray spectrum with time.\label{fig:HR}}
\end{figure}

\clearpage

\end{document}